\documentstyle[12pt]{article}
\title{Relativistic Quantum Measurements,\\
Unruh effect and Black Holes}
\author{
Michael B. Mensky\\[2mm]
P.N.Lebedev Physical Institute, 117924 Moscow, Russia
}

\date{}

\newcommand{\mathrm}{\rm}
\newcommand{\mathbf}{\bf}

\newcommand{\eq}[1]{(\ref{#1})}
\newcommand{\Fig}[1]{Fig.~\ref{#1}}
\newcommand{\be}{\begin{equation}}
\newcommand{\ee}{\end{equation}}
\newcommand{\ba}{\begin{eqnarray}}
\newcommand{\ea}{\end{eqnarray}}
\newcommand{\ban}{\begin{eqnarray*}}
\newcommand{\ean}{\end{eqnarray*}}
\newtheorem{remark}{Remark}

\renewcommand{\a}{{\mathbf a}}

\newcommand{\p}{{\mathbf p}}
\newcommand{\lC}{{\lambda_C}}
\newcommand{\Da}{\Delta a}
\newcommand{\x}{{\mathbf x}}
\renewcommand{\S}{{\cal S}}

\newcommand{\dmu}{\stackrel{\leftrightarrow}{\partial_{\mu}}}
\newcommand{\dnu}{\stackrel{\leftrightarrow}{\partial_{\nu}}}
\newcommand{\const}{{\mathrm{const}}}

\begin{document}
\maketitle
\abstract{It is shown how the technique of restricted path
integrals (RPI) or quantum corridors (QC) may be applied for the
analysis of relativistic measurements. Then this technique is
used to clarify the physical nature of thermal effects as seen by
an accelerated observer in Minkowski space-time (Unruh effect)
and by a far observer in the field of a black hole (Hawking
effect). The physical nature of the ``thermal atmosphere" around
the observer is analyzed in three cases: a)~the Unruh effect,
b)~an eternal (Kruskal) black hole and c)~a black hole forming in
the process of collapse. It is shown that thermal particles are
real only in the case (c). In the case (b) they cannot be
distinguished from real particles but they do not carry away mass
of the black hole until some of these particles are absorbed by
the far observer. In the case (a) thermal particles are virtual.}

\section{Introduction}\label{intro}

Nonrelativistic quantum theory of measurements is essentially based
on the von Neumann's postulate and cannot be applied for
relativistic systems because of the violation of causality in the
instantaneous state reduction of the measured system. This
problem was considered by many authors (see for example
\cite{Hellw}-\cite{Finkelst94}), but no consensus has been achieved
about how relativistic quantum measurements may be correctly
described. The general conclusion that may be drawn from this
discussion is that duration of a quantum measurement in time and
dimension of the area where the measurement is arranged 
cannot be neglected in the relativistic case. Relativistic 
quantum measurements must be considered as continuous both 
in space and time.

The restricted-path-integral (RPI) approach to continuous measurements
has been successfully applied to relativistic as well as
non-relativistic measurement setups
\cite{M79}-\cite{BorzM95}.
Particularly, this approach was used in \cite{BorzM95} to describe the
measurement of the position of a relativistic particle.

In what follows we shall elaborate this method in such a way that it
might be applied for a wide scope of quantum measurements on
elementary particles. Then some qualitative conclusions will be made
with the help of this technique for the Unruh and Hawking effects.

The RPI approach has been initiated by
R.Feynman \cite{Feynman48} to describe continuous
(prolonged in time) non-relativistic quantum measurements and was
technically elaborated and extended on new
areas in \cite{M79,book93,AudM97} (see also \cite{RPI-other}). An
important advantage of the approach is its being general and 
model-independent. 

The idea of the RPI
approach is that the evolution of the system undergoing a continuous
measurement must be described by the path integral restricted on the
set of paths compatible with the measurement readout. Therefore an
integral over a corridor of paths arises instead of the Feynman path
integral over all paths. This corridor of paths may be called {\em
quantum corridor} (QC) in analogy with the close (but different)
concept of the quantum trajectory introduced by H.Carmichael  \cite{Carmichael}. QCs play an
important role in the interpretation of continuous quantum
measurements. A certain set of QCs determines the continuous
measurement. Alternative QCs from this set correspond to alternative
measurement readouts possible in the given measurement.

In the present paper we shall outline some features of the method of
QCs for relativistic quantum particles. Then the concept of a QC will
be used to analyze some conceptual problems in connection with the
Unruh effect for an accelerated observer and the Hawking effect in
the field of a black hole. In the course of the analysis we shall
clarify the physical nature of the ``thermal atmosphere" observed by
an accelerated observer in Minkowski space-time or by an observer
moving far from a black hole. More concretely, we shall answer the
following questions:
\begin{itemize}
\item Is it possible, while observing thermal effects, to separate
contributions of different particles forming the thermal atmosphere?
\item Whether the particles constituting this atmosphere are real,
i.e. whether each of them may be observed in such a way that the fact
of its existence be independent of the measurement?
\end{itemize}
We shall see that the answers to these questions are different 
not only for the Unruh and Hawking effects, but also for the 
Hawking effect in the case of the``eternal" black hole (described 
by the Kruskal metric) and the black hole 
arising in the course of collapse. Some of the conclusions we shall
arrive at are of course known, particularly from the important paper
of W.Unruh and R.Wald \cite{UnruhWald}. However some of them, 
especially the difference between eternal black holes and those 
forming in collapse, seem to have never been formulated clearly 
enough.

\section{Relativistic Path Integrals}\label{path-relat}

The causal propagator (transition amplitude) for a relativistic
particle can be expressed in the form of a path integral if one
introduces, following E.C.Stueckelberg \cite{Stueck}, the fifth
parameter (besides four space-time coordinates) $\tau$ called
the {\em proper time} or {\em historical time}.

Consider for simplicity a scalar particle of the mass $m$. Its
{\em causal propagator} is equal to the integral over the
proper time,\footnote{We shall use in the present paper the
natural units $\hbar=c=1$.}
\be
K(x'',x')=\int_0^{\infty}d\tau\,
   \exp\left( -i(m^2-i\epsilon)\tau\right)
   \,K_{\tau}(x'',x'),
\label{tau-int}\ee
of a subsidiary {\em proper-time-de\-pen\-dent propagator}. The
latter, in turn, may be given the form of a path integral:
\be
K_{\tau}(x'',x')=\int_{x''\leftarrow x'} d[x]_{\tau}
  \, \exp\left( -\frac{i}{4}
  \int_0^{\tau}(\dot x,\dot x)d\tau\right).
\label{path-int}\ee
Here $(,)$ denotes the Lorentzian inner product and the path
$[x]_{\tau}$ between the points $x'$ and $x''$ of the Minkowski
space-time is parametrized by the interval of the proper time $[0,
\tau]$.\footnote{Notice that this proper time does not coincide with
what is called proper time in classical physics (the proper time of an
observer at the given trajectory). This is why the term `historical
time' seems more appropriate. However `proper time' is used in this
context more often.}

As a result of these definitions, the subsidiary
proper-time-dependent propagator satisfies the ``relativistic
Schr\"odinger-type equation"
\be
\frac{d}{d\tau}K_{\tau}(x'',x')
  = -i\, \Box K_{\tau}(x'',x')
\label{rel-Schroed}\ee
and the causal propagator $K(x'',x')$ is a Green function of
the Klein-Gordon equation:
\be
(\Box + m^2)K(x'',x') = -i\delta (x'',x').
\label{GreenFuncEq}\ee

Being a Green function, the propagator $K(x'',x')$ satisfies an
important relation
\be
i\int_{\S} \sigma^{\mu} \,K(x'',x) \dmu K(x,x')= K(x'',x')
\label{Kolmogor-close}\ee
where $\S$ is a closed hypersurface with the point $x'$ being inside
and $x''$ outside it,\footnote{An analogous relation but with the
opposite sign in the r.h.s. is valid also for $\S$ having $x''$ inside
and $x'$ outside it.} $\sigma^{\mu}$ is an element of area of the
hypersurface and $\dmu$ is defined by
$$
f(x)\dmu g(x)=
f(x)\stackrel{\leftrightarrow}{\frac{\partial}{\partial x^{\mu}}}g(x)=
f(x)\frac{\partial g(x)}{\partial x^{\mu}}
-\frac{\partial f(x)}{\partial x^{\mu}}g(x).
$$

These properties of the propagator may be generalized for the case of
an arbitrary electromagnetic or gravitational field. All the
derivatives must be covariant in this case and the path integral
\eq{path-int} should be defined covariantly \cite{CovarPI}.

For the analysis of continuous measurements on relativistic
particles in the framework of the RPI
approach we have to deal with path integrals of the type of
Eqs.~(\ref{tau-int},~\ref{path-int}) but restricted on the sets
of paths compatible with the corresponding measurement outputs.

\section{Measurements on particles}\label{SectDifProcesses}

Let us shortly consider the main features of the relativistic RPI
method starting with the simple case of the measurement of the
particle position.

\subsection{Measurement of the position}

\begin{figure}
\let\picnaturalsize=N
\def\picsize{6cm}
\def\picfilename{pairfig1.eps}
\ifx\nopictures Y\else{\ifx\epsfloaded Y\else\input epsf \fi
\let\epsfloaded=Y
\centerline{\ifx\picnaturalsize N\epsfxsize \picsize\fi \epsfbox{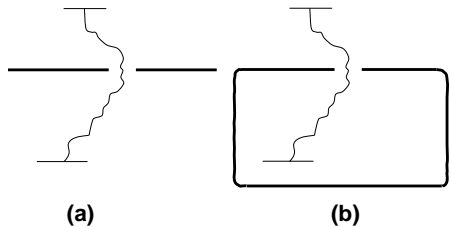}}}\fi
\caption{The measurement of the position of a relativistic particle may be presented by paths going through the gate in a time slice. Alternative measurement outputs are presented by different gates in the given time slice (a). In the general situation a closed hypersurface should be taken instead of a time slice (b). Time axis is directed upward in this as well as in the following figures.}
\label{Fig1}
\end{figure}

The measurement of the particle position at a time moment $x^0 = t$
resulting in the measurement output $\x = \a$ may be described
\cite{BorzM95} by the integral over paths intersecting the space-like
surface $\S=\{ x | x^0=t = \const\}$ in a narrow region around the
point $a=(t,\a)$ (see Fig.~\ref{Fig1}a). We shall say that the paths
go through the gate in the surface $\S$, the location of the gate
corresponding to the measurement output $\a$ and the width equal to
the measurement resolution $\Da$.

\begin{remark}
Actually the surface $\S$ must be closed as is shown in
Fig.~\ref{Fig1}b, in accord with the relation \eq{Kolmogor-close}. In
the case of null external field the integral over the past spacelike
hypersurface of $\S$ as well as the integrals over the timelike side
hypersurfaces are zero provided the side hypersurfaces are far enough.
Therefore $\S$ may in this special case be taken to be a time slice
$\{ x | x^0=t = \const\}$. In the present paper we shall consider the
general situation, hence closed surfaces with gates will play the main
role.
\end{remark}

Thus, the result of the measurement equal to $a\in\S$ 
may be described by a small area (gate) $G(a)$ around 
the point $a\in\S$ on the surface $\S$. The corresponding
amplitude $K^{G(a)}(x'',x')$, describing the evolution of a particle
undergoing the measurement under the condition that the measurement
gave the result $a$, should be defined as a path integral over the
paths going from $x'$ to $x''$ through the gate $G(a)$. The integral
is the product of two integral, one from $x'$ to $G(a)$ and the other
from $G(a)$ to $x''$. Each of these two integrals is close (though not
equal) to the complete propagator between the corresponding points.
This is the reason why the amplitude $K^{G(a)}(x'',x')$ may be defined
\cite{BorzM95} directly through these propagators:
$$
K^{G(a)}(x'',x')
=\int_{b\in G(a)} \sigma^{\mu}(b)\,K_{\mu}^{(b)}(x'',x')
$$
where
\be
K_{\mu}^{(b)}(x'',x')=
i\, K(x'',b)
\dmu(b)
K(b,x').
\label{precise-meas-ampl}\ee
This amplitude corresponds to the RPI in the corridor presented in
Fig.\ref{Fig1}. This corridor is a (closed) hypersurface with the
gate. We shall consider generalizations of this quantum corridor in
the following sections.

The amplitude (\ref{precise-meas-ampl}) is derived for the particle
which is in the space-time point $x'$ before the measurement and in
the point $x''$ after it. The realistic situation corresponds usually
to the initial and final states given by the wave functions at the
corresponding time moments $t'$, $t''$ (presented by short horizontal
lines in Fig.~\ref{Fig1}). The measurement amplitude
\eq{precise-meas-ampl} must then be multiplied by the corresponding
wave functions and integrated over time slices $t'$ and $t''$:
\be
K_{\kappa}^{(b)}(\psi'',\psi')
  =-\int \sigma^{\mu}(x'') \,\sigma^{\nu}(x') \,
  \overline{\psi''^(x'')}\,
\dmu(x'')
  \,K_{\kappa}^{(b)}(x'',x')\,
\dnu(x')
  \,\psi'(x')
\label{pos-wavef}\ee
(the bar denotes a complex conjugate).

The relation \eq{Kolmogor-close} (corresponding in the
non-relativistic case to conservation of probabilities or
unitarity of the evolution operator) may be shown to lead to the
``generalized unitarity" of the measurement amplitudes provided that
the dimension of $G(a)$ is larger than the Compton wavelength
$\lC=1/m$ of the measured particle. The physical reason is that the
localization of the particle in a region of the size $\Da$ requires
energy of the order of $1/\Da$ and may therefore lead to creation of
pairs if $\Da < \lC$. Such a pair creation is caused by no external
reason but is induced by the measurement itself. It distorts the
picture of what happens and is therefore a sort of ``measurement
noise". The condition $\Da > \lC$ guarantees that the measurement
noise is negligible and the observed particle is real.

The condition $\Da > \lC$ makes sense only for a massive particle.
However the more general condition $\Da > \lambda$ may be applied
for a massless particle. Here $\lambda=1/p$ is the ``typical"
wavelength of the particle in the conditions which the measurement
is performed in. It is determined by the ``typical" linear
momentum $p$. This condition guarantees that the localization
of the particle in the region of the dimension $\Da$ does not
result in the creation of pairs of particles having momenta
of the same order as the momentum of the measured particle. In this
case the measurement noise is small in the interval of momenta
(wavelengths) which is interesting. The particle observed in
this interval of momenta may be interpreted as real.

\subsection{Other relativistic effects}\label{SectOtherEf}

\begin{figure}
\let\picnaturalsize=N
\def\picsize{7cm}
\def\picfilename{pairfig2.eps}
\ifx\nopictures Y\else{\ifx\epsfloaded Y\else\input epsf \fi
\let\epsfloaded=Y
\centerline{\ifx\picnaturalsize N\epsfxsize \picsize\fi 
\epsfbox{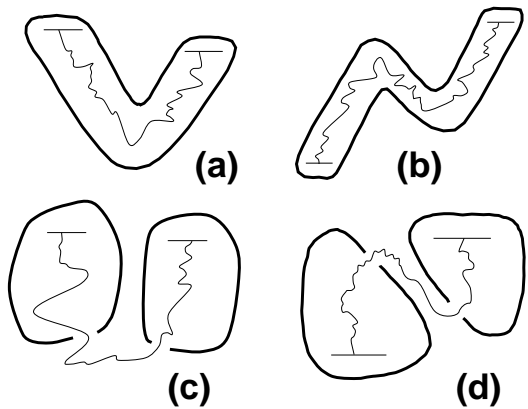}}}\fi
\caption{Relativistic measurements presented by corridors of paths 
(quantum corridors) in the space-time.  Pair creation is presented in 
(a, c) and a causal zig-zag in (b, d). The short horizontal lines 
correspond to the parts of the spacelike hypersurfaces on which in- 
and out-particles are registered.}
\label{Fig2}
\end{figure}

Many relativistic measurements (real or thoughtful experiments) may be
characterized in the framework of the RPI approach by corridors of
paths (quantum corridors) i.e. closed hypersurfaces, may be with gates
in them. Examples are given in Fig.~\ref{Fig2}.
Two alternative schemes are presented in Fig.~\ref{Fig2}~(a,c) for the
observation of the pair creation and in Fig.~\ref{Fig2}~(b,d) for the
observation of the causal zig-zag.

The process of measurement may result in a number of alternative
measurement outputs. If the measurement is described by quantum
corridors, different alternatives correspond to different corridors.
Thus, the V-type corridor in \Fig{Fig2}a is only one of many
alternative corridors with different locations of the vertex.
Analogously, the corridor of \Fig{Fig2}b is one of the corridors
describing propagation from one time slice to another one with the
trajectory observed (measured) with a finite resolution. For the
realization of both these types of measurement one needs a medium
consisting of objects (for example photons) weakly interacting with
the measured particle and thus localizing it,  with finite resolution,
in space and time.

Another type of measurement corresponds to the corridors with gates in
\Fig{Fig2}cd. In this case (just as for the position measurement,
\Fig{Fig1}), the closed surface is fixed and the alternative
measurement results correspond to different locations of the
gates.\footnote{In what follows we shall consider also such
measurement schemes that the number of gates may also be different for
different measurement results.} This formal scheme describes the 
observation arranged at the given closed surface with the width of the
gates presenting the resolution of the observation. Such a measurement
requires a net of objects activated in the specified time moments. We
do not need to specify details of this realization because the
method of quantum corridors does not depend on the concrete
measurement setup but only on the kind of information supplied by 
the measurement.

It is essential how wide is the corridor or the gate. To make this
question clear, it is reasonable to calculate restricted path
integrals (RPI) in the situation when there is no fields which could
cause non-trivial processes (for example pair creation or causal
zig-zag). One may expect that in this situation all RPI corresponding
to Fig.~\ref{Fig2} must have negligible values. This may be shown
valid if the corridor and the gate are wider than the Compton
wavelength of the measured particle, $\Da\gg\lC$ or, more generally,
if the corridor and the gate are wider than the typical wavelength of
the measured particle, $\Da\gg\lambda$.

If the width of the corridor or the gates is less (or of the order of)
the Compton length, then the result of the RPI calculation is non-zero
even for null fields. The physical reason of this fact is that the
localization of a particle in the region of smaller dimension than
$\lC$ requires inserting energy larger than the proper energy of the
particle. This energy may lead to the creation of pairs. In this case
pairs are created because of the too detailed observation of what
happens. Pair creation is then the effect of the measurement itself,
not of any external reason. If $\Da \gg\lC$, the measurement does not
induce pair creation. The observed particles are then real. The weaker
condition $\Da \gg\lambda$, where $\lambda$ is a typical wavelength of
the measured particle, guarantees that the particles cannot be created
with momenta of the order of one interesting for us. In this case the
observed particle is real provided that we are not interested in
momenta less than $1/\Da$.

Considering relativistic measurements in non-zero fields, we have
therefore to choose wide enough corridors and gates ($\Da > \lC$ or
$\Da > \lambda$) to avoid too strong influence of the measurement. If
on the contrary the influence of the measurement is the aim of the
investigation in its own right, then the width of the corridor or the
gate may be less than the wavelength.

\section{Unruh effect}

As it has been shown by W.Unruh \cite{Unruh76}, an accelerated
observer in Minkowski space-time will see the vacuum as a thermal bath
with the temperature proportional to its acceleration, $kT = w/2\pi$.
This phenomenon was called {\em Unruh effect}. It is convenient to
analyze the Unruh effect in the Rindler coordinates $(\eta,\xi)$,
which are related to Minkowski coordinates $(x^0,x^1)$ by the
transformation
\begin{equation}
  x^0 = {1 \over  w} e^{w\xi} \sinh w\eta , \qquad
  x^1 = {1 \over  w} e^{w\xi} \cosh w\eta.   \label{RindlerCoord}
\end{equation}
The trajectory of the accelerated observer has in the Rindler
coordinates the simple form $\xi = 0$ and the Rindler time $\eta$ is a
proper time on this trajectory. The surfaces $x^1 = \pm x^0$ are event
horizons for the accelerated observer. This means that only those
events may be causally connected with him which are in the same
quadrant in respect to the horizons. This quadrant is sometimes called
``Rindler wedge".

The causal propagator of the (massless) particle in the Minkowski
space-time between two points on the trajectory of the accelerated
observer is equal (up to the number factor) to \cite{TroostVanDam}
\be
\frac{w^2/4}{\sinh^2 \frac 12 w(\eta' - \eta'')} =
\sum_{n = - \infty}^{\infty}
\frac{1}{[(\eta' - \eta'') + i\beta n]^2}
\label{TroostProp}\ee
where $\beta = (kT)^{-1}$. In the energy-momentum representation
\be
\frac{w^2/4}{\sinh^2 \frac 12 w(\eta' - \eta'')} =
-\frac{1}{(2\pi)^2}\int dE\,d\p \; e^{iE(\eta' - \eta'')} D_{\beta}(E, \p)
\label{TroostEnergyRep}\ee
the propagator has the form
\be
D_{\beta}(E, \p)=
\frac{i}{E^2-\p^2 + i\epsilon} + \frac{2\pi\delta(E^2-\p^2)}{e^{\beta|E|}-1}
\label{TroostPropEn}\ee
with the first term corresponding to $n=0$ in \eq{TroostProp}.

This means that the usual propagator in the Minkowski space-time has
the form of the thermal propagator in respect to the Rindler time (a
proper time of the accelerated observer). Formally this leads to the
conclusion that the accelerated observer will see the thermal bath
instead of the vacuum. In the expansion (\ref{TroostProp}) the terms
with $n\ne 0$ are responsible for the thermal effects.

\begin{figure}
\let\picnaturalsize=N
\def\picsize{6cm}
\def\picfilename{pairfig3.eps}
\ifx\nopictures Y\else{\ifx\epsfloaded Y\else\input epsf \fi
\let\epsfloaded=Y
\centerline{\ifx\picnaturalsize N\epsfxsize \picsize\fi 
\epsfbox{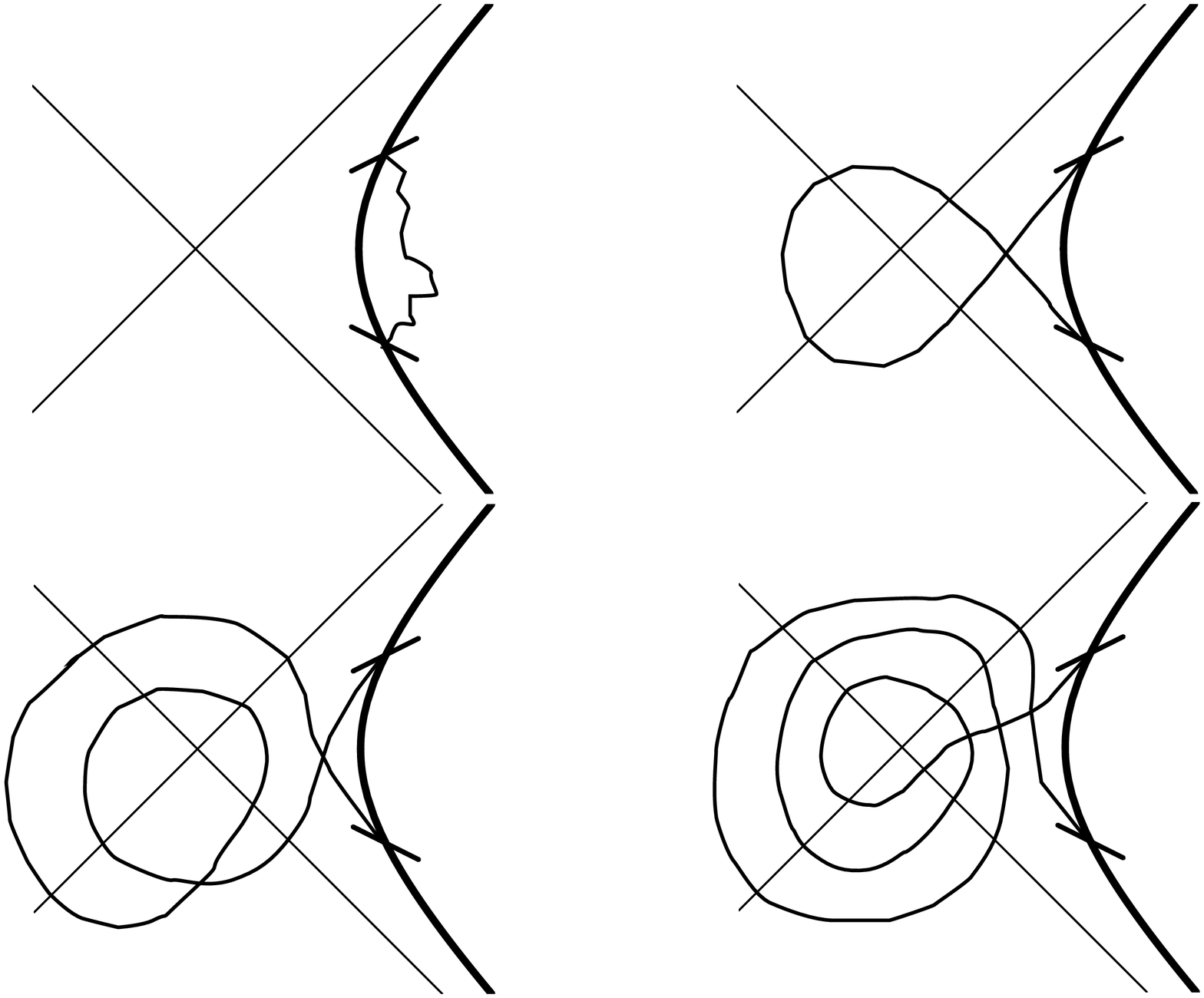}}}\fi
\caption{Paths with different winding numbers $n$ in respect to the 
origin of the Rindler plane. Topologically nontrivial paths $n\ne 0$ 
are responsible for the Unruh effect. The trajectory of the 
accelerated observer (thick line) and his event horizons (thin direct 
lines) are also drawn in the figure. }
\label{Fig3}
\end{figure}

As it is shown by W.Troost and H.Van Dam \cite{TroostVanDam}, 
in the path-integral
representation of the propagator {\em the term with the given $n$ is
presented by the paths having the winding number $n$} in respect to
the origin of the plane $(x^0, x^1)$. This means that only those paths
which go around the origin precisely $n$ times contribute to the $n$th
term (see Fig.~\ref{Fig3}).
Thermal effects are therefore presented by topologically non-trivial
paths having $n\ne 0$.

\begin{figure}
\let\picnaturalsize=N
\def\picsize{6cm}
\def\picfilename{pairfig4.eps}
\ifx\nopictures Y\else{\ifx\epsfloaded Y\else\input epsf \fi
\let\epsfloaded=Y
\centerline{\ifx\picnaturalsize N\epsfxsize \picsize\fi 
\epsfbox{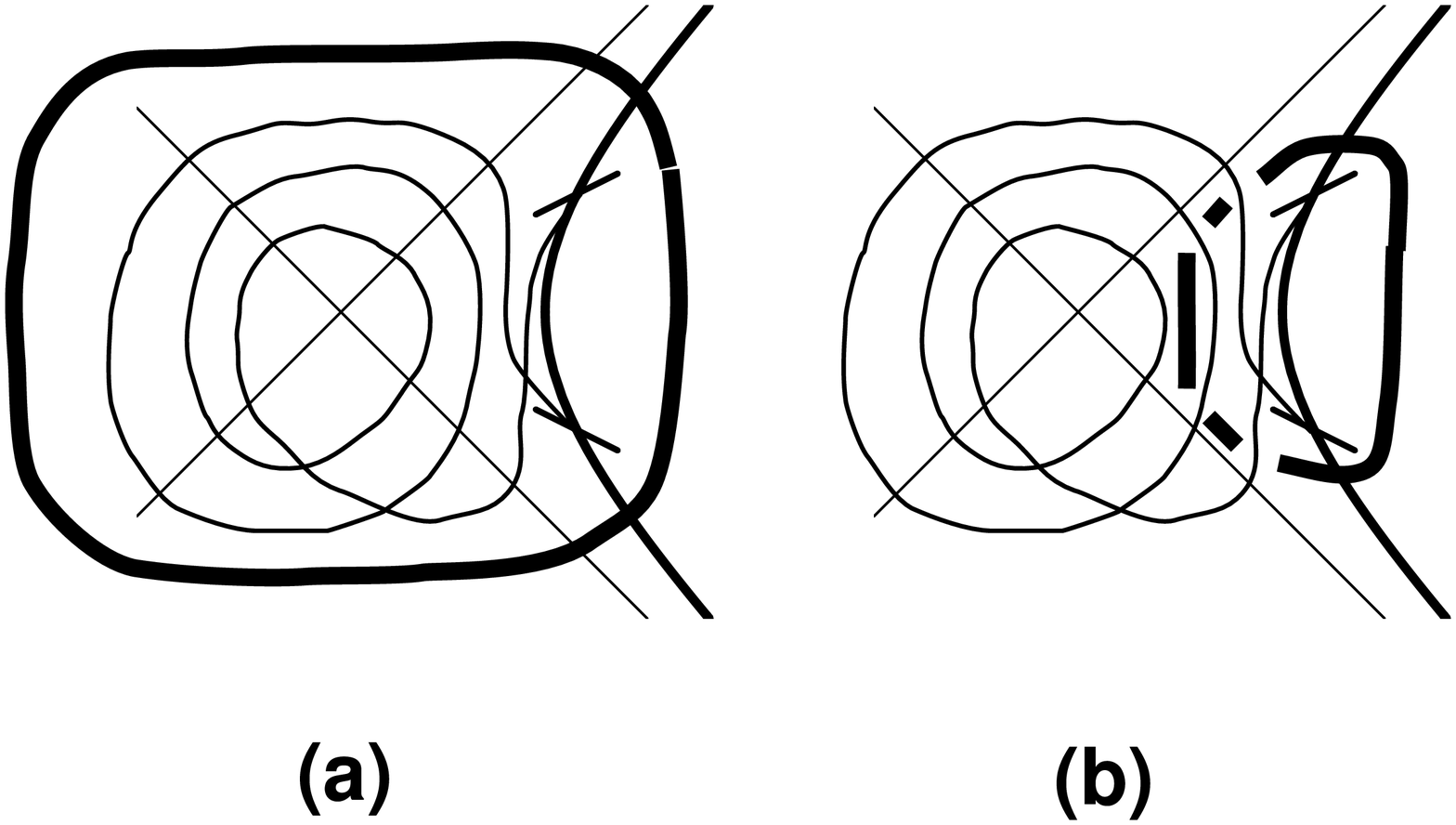}}}\fi
\caption{Observation of the Unruh effect: (a) the measurement 
includs the whole ``thermal effect" as the result of vacuum 
fluctuations (wide corridor); (b)  the observation distinguishes 
single thermal particles, however their creation under influence of 
the measurement cannot be excluded (narrow corridor).}
\label{Fig4}
\end{figure}

Let us apply now the RPI approach to analyze the Unruh effect.
Consider first the measurement setup which does not induce pair
creation.  According to what has been said at the end of
Sect.~\ref{SectOtherEf}, if we want to arrange the observation in such
a way that the measurement itself does not induce the pair creation,
then we have to choose the quantum corridors (describing this
measurement) wider than the wavelengths of particles in the given
thermal bath. The typical energy for these particles is $kT$, so that
the wavelength is of the order of $\lambda = 1/kT$. It can be shown
that any point in the trajectory of the accelerated observer is
separated by the distance of the order of $\lambda$ from the
corresponding point at the ``trajectory of the antiobserver" obtained
by the reflection through the origin ($x^0\rightarrow -x^0$,
$x^1\rightarrow -x^1$).  Therefore among all alternative wide
corridors we have to consider those which include, together with any
part of the observer trajectory, also the corresponding part of the
trajectory of the ``antiobserver" and the whole region between these
lines. All topologically nontrivial paths responsible for thermal
effects will be included in this corridor (Fig.~\ref{Fig4}a).
Individual particles from the ``thermal atmosphere" of the accelerated
observer cannot be separated with the help of the measurement of this type.
Thermal terms are interpreted in this case as ``vacuum fluctuations in
the Minkowski vacuum".

Consider now another type of measurements, for which all alternative
corridors are restricted by the event horizons of the accelerated
observer (are enclosed in the ``Rindler wedge"). Then these corridors
are narrow (as compared with the wavelength). Effects of the
measurement cannot be excluded in such a measurement. Let us analyze
this type of measurements.

It is possible to choose narrow corridors to characterize thermal
effects in more detail. For example the quantum corridor presented in
\Fig{Fig4}b gives an amplitude for the propagation of the particle
with the creation of not less than two thermal particles ($n\ge 2$),
one of which travels freely through the area of the measurement. It is
seen from \Fig{Fig4}b that the thermal effect will be interpreted by
the corresponding observer as the effect of particles coming from the
past horizon and going to the future horizon. Amplitudes corresponding
to the corridors of this type may be calculated  and in principle they
may be compare with experimental data. However the corridor in this
case will be narrow (as compared with the wavelength of thermal
particles). This means that the influence of the measuring setup is
not negligible in the corresponding experiments. The observed
particles cannot be interpreted as real particles existing
independently of the measurement.

\begin{figure}
\let\picnaturalsize=N
\def\picsize{6cm}
\def\picfilename{pairfig5.eps}
\ifx\nopictures Y\else{\ifx\epsfloaded Y\else\input epsf \fi
\let\epsfloaded=Y
\centerline{\ifx\picnaturalsize N\epsfxsize \picsize\fi \epsfbox{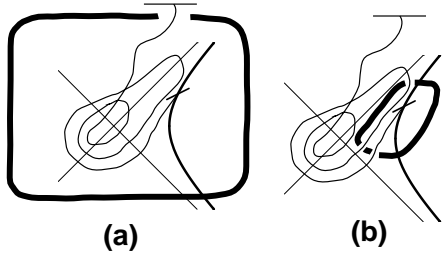}}}\fi
\caption{Absorption of a particle by an accelerated observer: 
(a)~Observation in a wide corridor of two real processes, 
absorption of a `Rindler particle' and creation of the corresponding `Minkowski particle'. Contributions of single thermal particles cannot 
be separated. (b)~Observation in a narrow corridor separates 
contributions of single thermal particles, but creation of these 
particles under the influence of the measurement cannot be excluded.}
\label{Fig5}
\end{figure}

We can consider path integrals describing the absorption of a thermal 
particle by the accelerated observer (\Fig{Fig5}).
The absorption of a ``Rindler particle" is accompanied in this case by
the creation of one more particle which can be absorbed by an inertial
observer. A wide corridor (with a wide gate) exists in this case
(\Fig{Fig5}a). It includes all thermal terms (all winding numbers
$n$). The absorption of a `Rindler particle' and accompanying
radiation of a `Minkowski particle' is a real (not virtual) process,
but the contributions of different $n$ to this process cannot be
separated experimentally. A more detailed observation separating these
contributions is described by narrow quantum corridors (\Fig{Fig5}b).
However the influence of the measuring setup is not negligible in 
this case. When observing thermal particles, the observer cannot
interpret them as real ones existing independently of the observation.

Conclusions for the Unruh effect:
\begin{itemize}
\item The ``thermal atmosphere" of an accelerated observer consists of
virtual rather than real particles which are parts of a long loop
presenting a vacuum fluctuation (\Fig{Fig4}a).
\item The observation performed in a narrow region (as compared
with the wavelengths of thermal particles) may lead to ``discovery" of 
single thermal particles, but the influence of the measurement onto 
creation of these particles cannot be excluded (\Fig{Fig4}b).
\item If a thermal particle is absorbed by an accelerated observer,
then the loop is broken and the counterpart antiparticle becomes real
and may be observed as a real particle (\Fig{Fig5}a).
\end{itemize}

\section{Black Holes}

\begin{figure}
\let\picnaturalsize=N
\def\picsize{6cm}
\def\picfilename{pairfig6.eps}
\ifx\nopictures Y\else{\ifx\epsfloaded Y\else\input epsf \fi
\let\epsfloaded=Y
\centerline{\ifx\picnaturalsize N\epsfxsize \picsize\fi \epsfbox{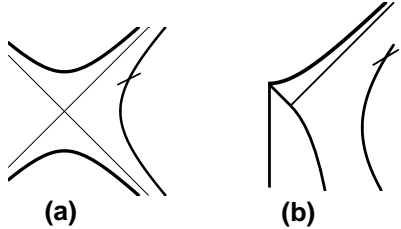}}}\fi
\caption{Two types of black holes: (a) an eternal (Kruskal) BH has
future and past event horizons (thin direct lines) and future and past
singularities (thick lines on the top and bottom of the diagram); (b)
a BH forming by collapse has only future singularity and only future
horizon starting at the surface of the collapsing body. The trajectory
of a far observer is presented in both cases (thick line on the right).}
\label{Fig6}
\end{figure}

Theoretically two qualitatively different types of black holes (BH)
may exist (see \Fig{Fig6})\footnote{We suppose that the reader is
familiar with basic features of BH which may be found for example in
\cite{Birrell82}}: an {\em eternal BH} and a {\em BH forming in the
process of collapse} of usual matter (for example a star). An eternal
BH (if it has null  angular momentum and charge) is described by the
Kruskal metric and have two event horizons (the future and past
horizons) and two singularities (the future and past singularities),
see \Fig{Fig6}a. A BH forming  in collapse have only one (future)
horizon and only one (future) singularity. For both types of BH a
trajectory of an observer moving at a constant distance from the BH is
drawn in \Fig{Fig6}.

It has been shown by S.Hawking \cite{Hawking} that an observer moving
far from the BH will see a thermal bath having the temperature
inversely proportional to the BH mass: $kT = 1/8\pi GM$ where G is the
gravitational constant. However the nature of thermal effects is not
quite clear up to now \cite{Unruh92,WaldBk94}. We shall apply the RPI
approach to analyze this question.

\subsection{An eternal black hole}

\begin{figure}
\let\picnaturalsize=N
\def\picsize{6cm}
\def\picfilename{pairfig7.eps}
\ifx\nopictures Y\else{\ifx\epsfloaded Y\else\input epsf \fi
\let\epsfloaded=Y
\centerline{\ifx\picnaturalsize N\epsfxsize \picsize\fi 
\epsfbox{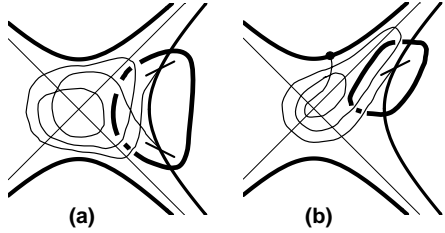}}}\fi
\caption{Observations in the field of an eternal BH: (a) The 
observation of the propagation arranged in a wide region 
simultaneously discovers singular thermal particles having 
all properties of real ones. No energy is taken 
from the BH. (b) The absorption of a particle observed in a wide 
region simultaneously discovers single thermal particles with the 
properties of real particles. The counterpart antiparticle is absorbed 
by the BH, extracting energy from it.}
\label{Fig7}
\end{figure}

As was demonstrated by W.Troost and H.Van Dam \cite{TroostVanDam}, thermal effects in the field of the eternal BH are 
(in complete analogy with the Unruh effect) described by 
the paths which are topologically non-trivial in respect to the
origin of the Kruskal coordinates (the point where the horizons cross
each other). Just as in the Rindler plane, the winding number in
respect to the origin of the Kruskal coordinates coincides with the
number of thermal particles. We shall analyze these paths with the
help of different quantum corridors (see Fig.~\ref{Fig7}).

Despite of the deep analogies, one feature essentially distinguishes the
Hawking effect from the Unruh effect. In the Unruh effect the
temperature tends to zero when the acceleration $w$ decreases (i.e.
for the observer far from the origin). In the Hawking effect the
temperature also decreases with the distance from the BH increasing.
However it tends to a constant value $1/8\pi GM$ for an infinitely far
observer. The temperature stays finite (and close to this constant) in
infinite interval of distances. Therefore, in the case of a BH wide
corridors around the observer trajectory do not include the origin 
of the Kruskal plane. We shall see that the existence of such 
corridors make possible the measurements separating the contributions 
of single thermal particles. 

Let us consider the measurement corresponding to the corridor of
\Fig{Fig7}a having the width larger than the typical wavelength of the 
thermal particles $\lambda = 1/kT$. Alternative measurement results 
are described in this case by the number and location of gates which 
are also wide enough. The number of gates (divided by 2) determines 
the number of thermal particles observed in the given measurement 
result. Since both the corridor and the gates are wide, all observed 
particles are real (not originated by the too narrow localization 
during the measurement). However, no particle is absorbed or issued in 
these processes by the BH, therefore the BH mass cannot be changed in 
this way. This is of course could be expected for the eternal BH.

The question naturally arises: if the observed thermal particle
cannot be distinguished from a real one, then it should carry an energy
and contribute to the general mass of the BH and its environment as it
is seen by a far observer. The answer is yes, each of these particles
contributes to the general energy, but the sum of all these
contributions is zero. This is connected with the special properties
of time of the far observer. A surface of constant time for such
an observer is presented at the Kruskal diagram (as in \Fig{Fig7}) by
the direct line passing through the origin. If the right end of such a
line goes upward (positive direction of time), its left end goes
downward.

Because of the loop-like structure of paths of thermal particles, they
are divided in pairs consisting of a particle and an antiparticle, the
particle in the `causal wedge of the observer' and the antiparticle in
the opposite wedge. In respect to the observer's time, the particle 
in each pair propagates in the positive direction of time, while
the corresponding antiparticle does in the negative time direction.
Therefore, if the particle have positive energy, the antiparticle has
negative energy. Because of the complete symmetry of the set of all
paths, these energies compensate each other so that the complete
contribution to the mass observed by the far observer is null.

The symmetry however breaks down if one of the particles is absorbed
by the observer (the corridor of \Fig{Fig7}b). The breakdown occurs 
at the moment of the absorption (in the observer's time). 
Beginning from this time moment the number of
antiparticles is larger by unit than the number of particles. The
absorption of  a thermal particle is accompanied by another process:
one of the antiparticles becomes real and falls onto the future
singularity of the BH. The negative energy of this antiparticle
contributes now to the general mass of the BH as it is seen by the 
observer.\footnote{Instead
of the future singularity, the free end of the torn loop may begin 
at the past singularity. Then the absorbed particle is issued by 
the BH.} The observer when measuring gravitational field will 
see that this field corresponds now to smaller general mass. 
This may be interpreted as diminishing of the mass of the BH.
From another point of view the origin of this mass is not the BH, 
but a particle moving in its vicinity.

Conclusions for the eternal (Kruskal) BH:
\begin{itemize}
\item Particles forming the ``thermal atmosphere" of the far observer
cannot be distinguished from real ones, but they do not change the
mass of the BH if they are not absorbed. Together with their
counterpart antiparticles they form a loop, and their energies (in
respect to the time of the far observer) compensate each other.
\item The absorption of a particle by the far observer is accompanied
by falling an antiparticle onto the singularity resulting in the
change of the BH mass as it is seen by the far observer. Instead, the
absorbed particle may be issued from the past singularity.
\end{itemize}

\subsection{A black hole forming in collapse}

\begin{figure}
\let\picnaturalsize=N
\def\picsize{7cm}
\def\picfilename{pairfig8.eps}
\ifx\nopictures Y\else{\ifx\epsfloaded Y\else\input epsf \fi
\let\epsfloaded=Y
\centerline{\ifx\picnaturalsize N\epsfxsize \picsize\fi 
\epsfbox{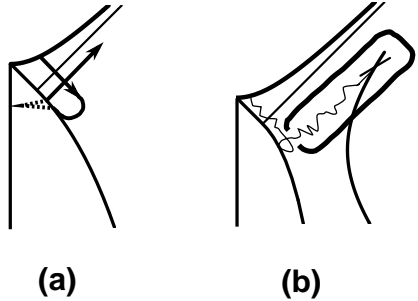}}}\fi
\caption{The BH forming in the process of collapse. (a) A virtual pair 
is converted in a real one at the point where the horizon is formed. 
(b) The particle absorbed by the observer is real and may be traced 
back to its origin.}
\label{Fig8}
\end{figure}

Consider now a BH forming in real collapse (\Fig{Fig8}). There is one
essential new feature of such a BH as compared with the eternal BH. In
the space-time point coinciding with the origin of the horizon, a
virtual pair may be ``torn off" with forming a real particle
escaping to infinity and a real antiparticle falling into the BH
(\Fig{Fig8}a). The particle of a virtual pair may go through the
collapsing body (which has at this stage the size of the order of the
wavelength of the considered particle), exit from the other side of
it and escape to infinity. The corresponding antiparticle bypasses the
body and falls into the BH.

As it is proved by S.Hawking \cite{Hawking}, the energy 
spectrum of particles
escaping to infinity due to this mechanism is thermal with the
temperature $kT = 1/8\pi GM$. This seems similar to what takes place
around the eternal BH (though due to another mechanism). However in
the case of the BH forming in collapse the ``thermal atmosphere"
consists of real rather than virtual particles. Creation of each
particle is accompanied by the creation of an antiparticle carrying
negative energy into the BH. The mass of the BH decreases in the
result of such a process. Each of the thermal particles formed in this
way may be absorbed by the far observer. The origin of the observed
particle may be in principle traced back to the moment of its
formation near the horizon origin (\Fig{Fig8}b).

Conclusions for the BH resulting in the process of collapse :
\begin{itemize}
\item Near the origin of the horizon a virtual pair may be converted
to a real one, with the antiparticle falling into the BH diminishing
its mass and the particle escaping to infinity along the horizon.
\item The real particles escaping to infinity may be seen (and
absorbed) by a far observer. However independently of their absorption
these particles are carrying mass of the BH away.
\item The trajectory of an absorbed (or only observed) particle may in
principle be traced back to the point near the origin of the horizon.
\end{itemize}

\section{Concluding remarks}

The technique of relativistic restricted path integrals (RPI) and
quantum corridors (QC) has been here only outlined. Some important
procedures characteristic for this technique were not properly
discussed, for example summing up over all alternative measurement
results. Besides, no RPI has been really calculated in the present
paper. Nevertheless, the estimate of the width of a QC in different
physical situations led us to some new conclusions or at least made
more clear some points of view on the Unruh and Hawking effects.

The main of these conclusions is a subtle distinction between the
cases of a)~the Unruh effect,  b)~an eternal black hole (BH) and
c)~the BH resulting in the process of collapse. The nature of the 
``thermal atmosphere" of the observer is different in these three 
cases. This atmosphere consists of virtual particles in the case (a) 
and of real particles in the case (c). In the intermediate case (b) 
thermal particles may be observed in a wide enough region so that they 
have all properties of real particles. If some of them are absorbed, 
the mass of the BH decreases by the corresponding amount. However 
until being absorbed these particles do not carry away mass of the BH 
so that this mass is constant.\footnote{The distinction between an 
eternal BH and one forming in collapse was discussed in \cite{Men76}. 
The conclusion was that the vacuum must be stable in the field of the 
eternal BH but not in the case of a collapsing body.} In fact, thermal 
particles attain more features of real particles with each step of 
advancing along the chain (a)$\rightarrow$(b)$\rightarrow$(c).

The difference between the eternal BH and the BH forming in collapse 
may have consequences for astrophysical observations. If some BH is 
observed, it is not necessary to expect that it will finally 
evaporate. This depends on its prehistory. If the BH has been formed 
by collapse, it will finally evaporate, but if it was existing at any 
time in the past, it will be existing infinitely also in the future. 
Such a BH is actually ``eternal". At least this is the case if the 
environment of the BH is not too dense, because otherwise absorption 
of particles from the ``thermal atmosphere" by the environment will 
lead to falling their antiparticle counterparts into the BH and 
resulting decrement of the observed BH mass.

\vspace{0.5cm}
\centerline{\bf ACKNOWLEDGEMENT}

This work was supported in part by the Russian Foundation for Basic
Research, grant 95-01-00111a.

\end{document}